\documentclass[showpacs]{revtex4}

\usepackage{dcolumn}
\usepackage{amsmath}
\usepackage[latin1]{inputenc}
\usepackage{graphicx, psfrag}
\usepackage{amssymb,gensymb}
\usepackage[colorlinks=true, citecolor=blue, urlcolor = blue, linkcolor= red, bookmarks=true]{hyperref}
\usepackage{float}
\usepackage{amsfonts}
\usepackage{dcolumn}
\usepackage{hyperref}
\usepackage{subfigure}
\usepackage{pgfplots}
\usepackage{epstopdf}
\usepackage{booktabs}
\usepackage{mathtools,mathrsfs}
\usepackage[english]{babel}

\begin{document}
	
\title[]{ Gravitational lensing by black holes in the $4D$ Einstein-Gauss-Bonnet gravity}
\author{Shafqat Ul Islam$^{a}$}\email{Shafphy@gmail.com}
\author{Rahul Kumar$^{a}$}\email{rahul.phy3@gmail.com}
\author{Sushant~G.~Ghosh$^{a,\;b}$} \email{sghosh2@jmi.ac.in}
	
\affiliation{$^{a}$ Centre for Theoretical Physics, Jamia Millia Islamia, New Delhi 110025, India}
\affiliation{$^{b}$ Astrophysics and Cosmology Research Unit, School of Mathematics, Statistics and Computer Science, University of	KwaZulu-Natal, Private Bag 54001, Durban 4000, South Africa}
\date{\today}

\begin{abstract}
Recently, a non-trivial $4D$ Einstein-Gauss-Bonnet (EGB) theory of gravity, by rescaling the GB coupling parameter as $\alpha/(D-4)$, was formulated in \cite{Glavan:2019inb}, which bypasses Lovelock's theorem and avoids Ostrogradsky instability. The theory admits a static spherically symmetric black hole, unlike $5D$ EGB or general relativity counterpart, which can have both Cauchy and event horizons. We generalize previous work, on gravitational lensing by a Schwarzschild black hole, in the strong and weak deflection limits to the $4D$ EGB black holes to calculate the deflection coefficients $\bar{a}$ and $\bar{b}$, while former increases and later decrease with increasing $\alpha$. We also find that the deflection angle $\alpha_D$, angular position $\theta_{\infty}$ and $u_{m}$ decreases, but angular separation $s$ increases with $\alpha$.
The effect of the GB coupling parameter $\alpha$ on positions and magnification of the source relativistic images is discussed in the context of SgrA* and M87* black holes. A brief description of the weak gravitational lensing using the Gauss-Bonnet theorem is presented.
\end{abstract}

\maketitle

\section{Introduction}
String theories, in the low-energy limits, give rise to the effective field theories of gravity such that the Lagrangian for these theories contains terms of quadratic and higher orders in the curvature in addition to the usual scalar curvature term \cite{Gross,Duff:1986pq,Gross2,Bento}.
Further, the gravitational action may be modified to include the quadratic curvature correction terms while keeping the equations of motion to second order, provided the quadratic terms appear in specific combinations corresponding to the Gauss-Bonnet (GB) invariants defined by \cite{Lanczos:1938sf}
\begin{eqnarray}\label{GB}
\mathscr{G} &=& R^2 -4 R_{\mu \nu}R^{\mu \nu}+R_{\mu \nu \rho \sigma}R^{\mu \nu \rho \sigma},
\end{eqnarray}
and such a theory is termed as $D\geq 5$ dimensional Einstein Gauss-Bonnet (EGB) gravity theory with the $D-4$ extra dimensions. It turns out that the four-dimensional GB term arises as the next to leading order correction to the gravitational effective action in string theory \cite{Gross,Duff:1986pq,Gross2,Bento}. The EGB gravity is a special case of the Lovelock theory of gravitation \cite{Lovelock:1971yv}. Since its equations of motion have no more than two derivatives of the metric, it has been proven to be free of ghosts \cite{Boulware:1985wk}.  Boulware and Deser \cite{Boulware:1985wk}, independently by Wheeler \cite{Wheeler:1985nh}, found exact black hole solutions in EGB gravity theories, which are generalizations of the Schwarzschild-Tangherlini  black holes \cite{Tangherlini:1963bw}.  However, the GB term  (\ref{GB})  is a topological invariant  in $4D$ as its contribution to all components of Einstein's equations are in fact proportional to $(D-4)$, i.e, it does not contributes to the equations of motion and one requires $D\geq 5$ for non-trivial gravitational dynamics. 

However, it was shown by Glavan and Lin \cite{Glavan:2019inb} that  by re-scaling the GB coupling constant, in the EGB gravity action,  as $\alpha\to \alpha/(D-4)$, the GB invariant makes a non-trivial contribution to the gravitational dynamics even in the $D=4$. This is evident by considering maximally symmetric spacetimes with curvature scale ${\cal K}$
\begin{equation}\label{gbc}
\frac{g_{\mu\sigma}}{\sqrt{-g}} \frac{\delta \mathscr{G}}{\delta g_{\nu\sigma}} = \frac{\alpha (D-2) (D-3)}{2(D-1)} {\cal K}^2 \delta_{\mu}^{\nu},
\end{equation}
obviously the variation of the GB action does not vanish in $D=4$ because of the re-scaled coupling constant \cite{Glavan:2019inb}. This $4D$  EGB gravity has already attracted much attentions and being extensively studied \cite{Konoplya:2020juj,Zhang:2020qew,Hegde:2020xlv,Zhang:2020qam,Lu:2020iav,Churilova}. The Boulware and Deser \cite{Boulware:1985wk} version of the spherically symmetric static  black hole  to the $4D$ EGB gravity was obtained in  \cite{Glavan:2019inb}, which was also extended to the charged case \cite{Fernandes:2020rpa}; this kind of solution for the Lovelock gravity has been obtained in Ref.~\cite{Konoplya:2020qqh,Casalino:2020kbt}. The corresponding black holes in a string cloud model was considered in \cite{Singh:2020nwo}. Also, nonstatic Vaidya-like spherical radiating black hole solutions have been obtained in $4D$ EGB gravity \cite{Ghosh1:2020vpc,Ghosh:2020syx}. More black hole solutions can be found in Refs.~\cite{Wei:2020ght,Kumar:2020owy,Doneva:2020ped,Konoplya:2020ibi,Kumar:2020uyz,Singh:2020xju,HosseiniMansoori:2020yfj}. Study of photon geodesics and the effects of GB coupling parameter on the shadow of $4D$ EGB black hole is presented in Ref.~\cite{Guo:2020zmf,Konoplya:2020bxa,Wei:2020ght,Kumar:2020owy}. 

The idea of $4D$ regularization of EGB gravity was originally initiated by Tomozawa \cite{Tomozawa:2011gp}, and later  Cognola {\it et al.} \cite{Cognola:2013fva} reformulated it by accounting quantum corrections due to a GB invariant within a classical Lagrangian approach. Though the Glavan and Lin \cite{Glavan:2019inb} $4D$ regularization procedure is currently a subject of dispute \cite{Ai:2020peo,Hennigar:2020lsl,Shu:2020cjw,Gurses:2020ofy,Mahapatra:2020rds}, however, several alternate regularization procedures have been proposed \cite{Lu:2020iav,Kobayashi:2020wqy,Hennigar:2020lsl,Casalino:2020kbt,Ma:2020ufk,Arrechea:2020evj,Fernandes:2020nbq}. Interestingly, neither of the critics \cite{Ai:2020peo,Gurses:2020ofy,Hennigar:2020lsl} disprove dimensional regularization procedure of \cite{Glavan:2019inb} at least for the case of spherically symmetric spacetimes. Furthermore, the static spherically symmetric $4D$  black hole  
solution found in \cite{Glavan:2019inb,Cognola:2013fva} remains valid for these alternate regularized theories \cite{Lu:2020iav,Hennigar:2020lsl,Casalino:2020kbt,Ma:2020ufk}. Furthermore, the semi-classical Einstein equations with conformal anomaly proportional to Euler density $\mathscr{G}$ \cite{Cai:2009ua,Cai:2014jea} and the $4D$ non-relativistic Horava-Lifshitz theory of gravity \cite{Kehagias:2009is} also admit the identical spherically symmetric black hole solution.

Gravitational lensing is one of the most powerful astrophysical tools for investigations of the strong-field features of gravity. Motivated by the pioneering work of Darwin \cite{Darwin}, strong gravitational lensing by compact astrophysical objects with a photon sphere, such as black holes, has been extensively studied.  The deflection of electromagnetic radiation in a gravitational field is commonly referred to as the gravitational lensing and an object causing a deflection is known as a gravitational lens. Virbhadra \cite{Virbhadra:1999nm} studied the strong-field situation to obtain the gravitational lens equation using an asymptotically flat background metric and also analyzed the gravitational lensing by a Schwarzschild black hole. The results have been applied to the supermassive black hole Sgr A* using numerical techniques \cite{Virbhadra:1999nm}. Later on, using a different formulation, an exact lens equation and integral expressions for its solutions were obtained \cite{Frittelli:1999yf}. However, it was Bozza {\it et al.} \cite{Bozza:2001xd} who first defined the strong-field deflection limit to analytically investigate the Schwarzschild black hole lensing. This technique has been applied to static, spherically symmetric metrics which includes  Reissner$-$Nordstrom black holes \cite{Eiroa:2003jf}, braneworld black holes \cite{Eiroa:2004gh,Whisker:2004gq,Eiroa:2005vd,Li:2015vqa}, charged black hole of heterotic string theory \cite{Bhadra:2003zs}, and was also generalized to an arbitrary static, spherically symmetric metric by Bozza \cite{Bozza:2002zj}. On the other hand, lensing in the strong gravitational field is a powerful tool to test the nature of compact objects, therefore, it continues to receive significant attention, and  more recent works include lensing from other black holes \cite{Chen:2009eu,Sarkar:2006ry,Javed:2019qyg,Shaikh:2019itn}. However, the qualitative features in lensing by these black holes, in the presence of a photon sphere, is similar to the Schwarzschild case. Also, strong gravitational lensing from various modifications of Schwarzschild geometry in modeling the galactic center has been studied, e.g., lensing from regular black holes was studied in \cite{Eiroa:2010wm,Ovgun:2019wej}, massive gravity black holes \cite{Panpanich:2019mll} and also lensing by wormholes \cite{Bronnikov:2018nub,Shaikh:2018oul}. 

The aim of this paper is to apply the prescription of Bozza {\it et al.} \cite{Bozza:2002zj} to investigate the gravitational lensing properties of the $4D$ EGB black hole. In particular, we calculate the strong lensing coefficients of the static spherically symmetric EGB black hole, from which we calculate the positions and magnification of the source's images and numerically compute them for the Schwarzschild black hole. The effect of the GB coupling parameter on the weak gravitational lensing is also investigated. The paper is organized as follows: we begin in Sect.~\ref{sec2} with a discussion of the static spherically symmetric black hole of the $4D$ EGB gravity. Gravitational deflection of light in strong-field limits of these black holes is investigated in Sect.~\ref{sec3}. Corresponding lensing observables and numerical estimations of deflection angle, image positions, and magnifications in the context of supermassive black holes, namely Sgr A* and M87* are presented in Sect.~\ref{sec4}. Section~\ref{sec5} is devoted to the weak gravitational lensing. Finally, paper ends in  Sect.~\ref{sec6} by summarizing our main findings. 

\section{The $4D$ EGB black hole}\label{sec2}
The EGB gravity action, with rescaled GB coupling $\alpha/(D-4)$ to restore the dimensional regularization, in the $D$-dimension spacetime  yields \cite{Glavan:2019inb}
\begin{eqnarray}
S &=& \frac{1}{16\pi G} \int d^{D}x\sqrt{-g}[R + \frac{\alpha}{D-4} \mathscr{G}],
\end{eqnarray}  
where $g$ is the determinant of metric tensor $g_{\mu\nu}$, $R$ is the Ricci scalar,  and $\alpha $ is the GB coupling constant considered to be positive. 
The $4D$ EGB theory is defined in the limit of $D\rightarrow 4$ at the level of equations of motion rather than in action, thereby the GB term makes a non-trivial contribution in gravitational dynamic \cite{Glavan:2019inb}, which admit static spherically symmetric black hole \cite{Glavan:2019inb} given by 
\begin{eqnarray}
ds^2 &=& -f(r) dt^2 + \frac{1}{f(r)}  dr^2 + r^2 (d\theta ^2 +\sin ^2\theta\,d\phi^2),\label{NR}
\end{eqnarray}
with
\begin{equation}
f_{\pm}(r)= 1+\frac{r^2}{32\pi \alpha G}\left( 1\pm \sqrt{1+\frac{128 \pi \alpha M G^2}{r^3}}\right).\label{fr}
\end{equation}
Here, $M$ is the black hole mass parameter and $G$ is the Newton's gravitational constant which is set to unity hereafter. Since various gravity theories \cite{Tomozawa:2011gp,Cai:2009ua,Cognola:2013fva,Casalino:2020kbt,Kehagias:2009is,Konoplya:2020qqh,Hennigar:2020lsl,Cai:2014jea,Ma:2020ufk} admit the spherically symmetric black hole  solution identical to Eq.~(\ref{NR}) with (\ref{fr}), therefore the presented study of gravitational lensing is also valid for these gravity black holes. It is clear that Eq.~(\ref{fr}) leads to two different branches of solutions corresponding to the $``\pm"$ sign. However, at large distances, metric function (\ref{fr}) reduces to 
\begin{equation}
f_-(r)=1-\frac{2M}{r},\;\;\;
f_+(r)=1+\frac{2M}{r}+\frac{r^2}{16\pi\alpha },
\end{equation}
and only $f_{-}(r)$ correctly identify with the Schwarzschild solution, even in the limit of vanishing GB coupling constant, $\alpha\to 0$, the metric (\ref{NR}) only with $f_{-}(r)$ in Eq.~(\ref{fr}) retains to the Schwarzschild black hole metric  \cite{Glavan:2019inb}. Henceforth, we shall restrict our discussion to $-$ve branch. 
 
To initiate discussion on gravitation lensing, we note that the static and spherically symmetric metric (\ref{NR}) is asymptotically flat and rewrite it  using the dimensionless variables as follow \cite{Bozza:2002zj}
\begin{eqnarray}
d\tilde{s}^2 &=& (2M)^{-2} ds^2 = -A(x)dT^2+\frac{1}{A(x)}dx^2+C(x)(d\theta ^2 +\sin ^2\theta\,d\phi^2),\label{NR1}
\end{eqnarray} 
where
\begin{eqnarray}
x = \frac{r}{2M},  \quad   T=\frac{t}{2M}, \quad \tilde{\alpha} = \frac{\alpha}{M^2},
\end{eqnarray} 
and accordingly from the metric (\ref{NR}), we have 
\begin{equation}
A(x) = 1+\frac{x^2}{8\pi \tilde{\alpha} }\left( 1 - \sqrt{1+\frac{16 \pi \tilde{\alpha}}{x^3}} \right), \quad C(x)=x^2.
\end{equation}
\begin{figure}
     \begin{tabular}{c c}	
		\includegraphics[scale=0.75]{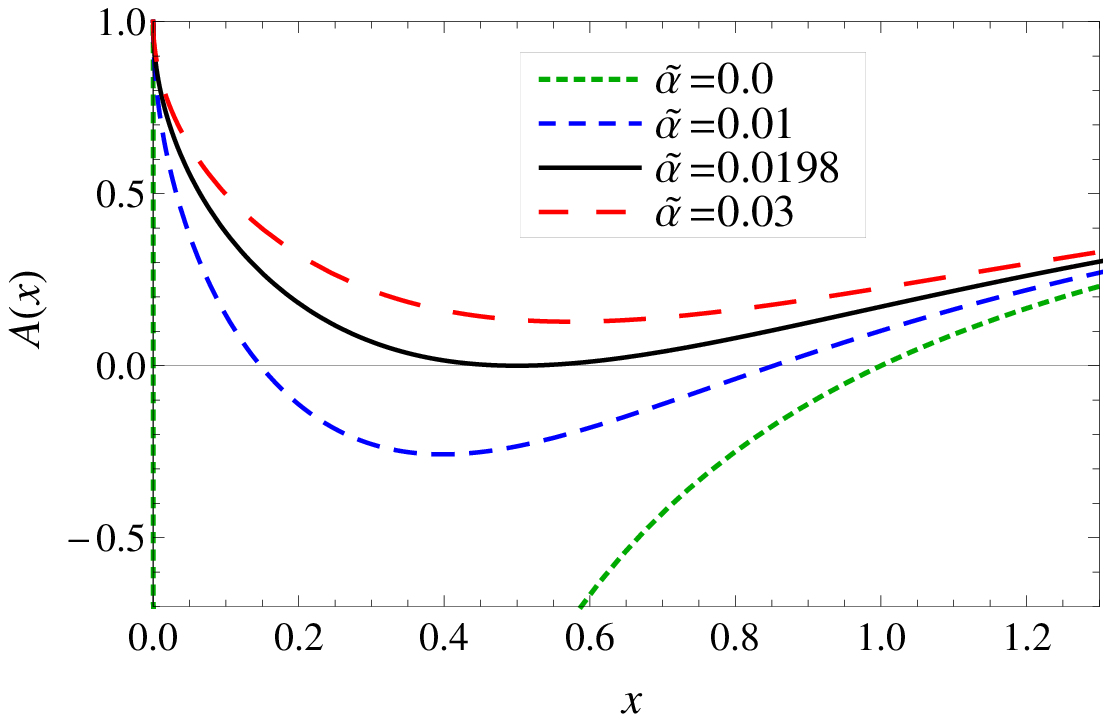}&
		\includegraphics[scale=0.75]{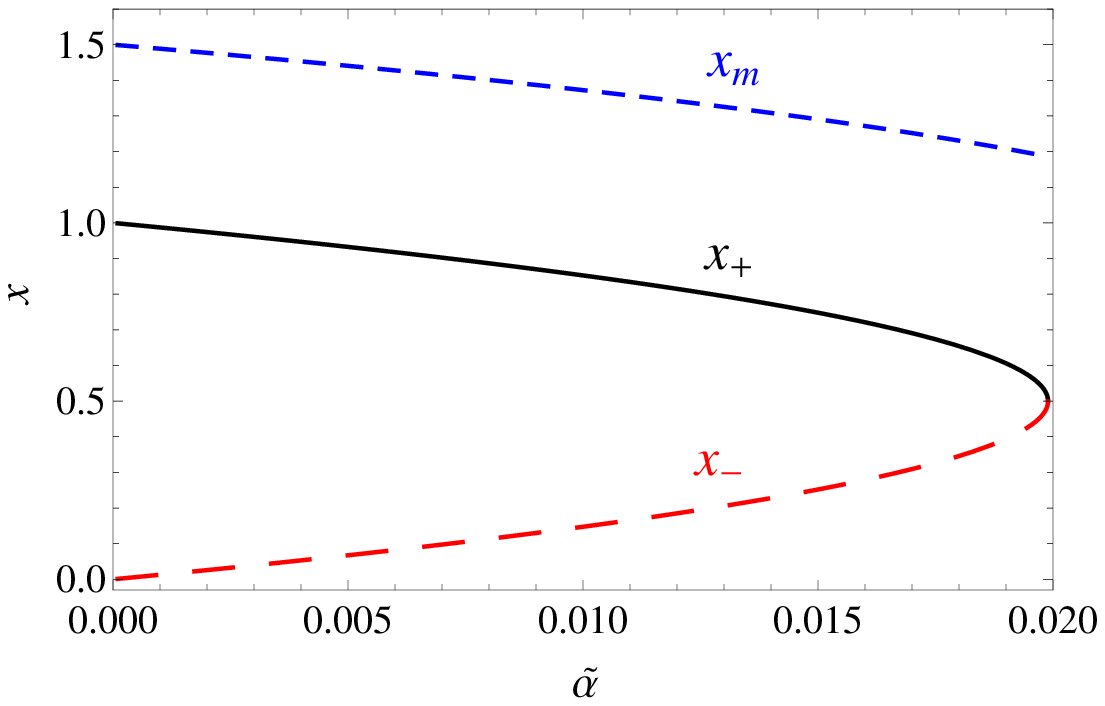}
	\end{tabular}
	\caption{(Left panel) Plot showing the horizons for various values of GB coupling parameter. The black solid line corresponds to the extremal value of $\tilde{\alpha}$. (Right panel) The behavior of event horizon radii $x_+$ (solid black line), Cauchy horizon radii $x_-$ (dashed red line), and photon sphere radii $x_m$ (dashed blue line) with GB coupling parameter $\tilde{\alpha}$.  }\label{plot}		
\end{figure} 
The event horizon is the largest positive root of  $g^{rr}=0$, i.e., horizon radii can be determined by solving 
\begin{equation}\label{Delta}
A(x)=0,
\end{equation}
which admits solutions
\begin{eqnarray}
x_{\pm} &=& \frac{1}{2}\Big(1 \pm \sqrt{1-16\pi \tilde{\alpha}} \Big).
\end{eqnarray}
Depending on the values of $\tilde{\alpha}$ and $M$, Eq.~(\ref{Delta}) has upto two real positive roots corresponding to the inner Cauchy horizon ($x_{-}$) and outer event horizon ($x_{+}$).  It turns out that $A(x)=0$ has no solution if  $\tilde{\alpha}>\tilde{\alpha}_E$ i.e., no black hole exists. Whereas, it has one double zero
if $\tilde{\alpha}=\tilde{\alpha}_E$, and two simple zeros if $\tilde{\alpha}<\tilde{\alpha}_E$, respectively, corresponding to the $4D$  EGB black hole with degenerate horizon ($x_{-}=x_{+}\equiv x_E$), and a  non-extremal black hole with two horizons ($x_{-}\neq x_{+}$) (cf. Fig.~\ref{plot}). The event horizon radii decrease whereas Cauchy horizon radii increase with increasing $\tilde{\alpha}$ (cf. Fig.~\ref{plot}). It is evident that the event horizon radii for the EGB black holes is smaller than the Schwarzschild black hole value. 

\section{Strong gravitational lensing}\label{sec3}
In this section, we focus on the gravitational deflection of light in the static spherically symmetric $4D$ EGB black hole spacetime (\ref{NR1}) and consider the propagation of light on the equatorial plane $(\theta =\pi/2)$, as due to  spherical symmetry, the same results can be applied to all $\theta$.  Then the metric (\ref{NR1}) reduces to 
\begin{eqnarray}\label{metric2}
d\tilde{s}^2 &=& -A(x)dT^2 + A(x)^{-1}dx^2 + C(x)d\phi ^{2}.
\end{eqnarray}
Since the spacetime is static and spherically symmetric, the projection of photon four-momentum along the Killing vectors of isometries are conserved quantities, namely the energy $\mathcal{E}=-p_{\mu}\xi^{\mu}_{(t)}$ and angular momentum $\mathcal{L}=p_{\mu}\xi^{\mu}_{(\phi)}$ are constant along the geodesics, where $\xi^{\mu}_{(t)}$ and $\xi^{\mu}_{(\phi)}$ are, respectively, the Killing vectors due to time-translational and rotational invariance \cite{Chandrasekhar:1992}. This yields
\begin{equation}
\frac{dt}{d\tau}=-\frac{\mathcal{E}}{A(x)},\qquad \frac{d\phi}{d\tau}=\frac{\mathcal{L}}{C(x)},\label{const}
\end{equation}
where $\tau$ is the afine parameter along the geodesics. Using Eq.~(\ref{const}) for the null geodesics equation $ds^2=0$, we obtain
\begin{equation}
\left(\frac{dx}{d\tau}\right)^2\equiv \dot{x}^2={\cal E}^2-\frac{\mathcal{L}^2A(x)}{C(x)}.
\end{equation}
The radial effective potential $V_{\text{eff}}(x)=\mathcal{L}^2A(x)/C(x)$, reads as
\begin{eqnarray}
V_{\text{eff}}(x) &=& \frac{\mathcal{L}^2}{x^2}\left(1+\frac{x^2}{8\pi \tilde{\alpha} }\left( 1 - \sqrt{1+\frac{16 \pi \tilde{\alpha}}{x^3}} \right)\right),
\end{eqnarray}
and describes the different kinds of possible orbits. In particular, photons simply move on circular orbits of constant radius $x_m$, at points where the potential is flat
\begin{equation}\label{co}
\frac{dV_{\text{eff}}(x)}{dx}=0\quad \Rightarrow\quad  \frac{A'(x)}{A(x)}= \frac{C'(x)}{C(x)},
\end{equation}
Eq.~(\ref{co}) reduces to 
\begin{equation}
4 x^3 - 9 x + 64 \pi \tilde{\alpha}  = 0,
\end{equation}
and solving which, gives the radius of photon circular orbits $x_m$; at $x=x_m$ potential has a unique maximum. These orbits are unstable against small radial perturbations, which would finally drive photons into the black hole or toward spatial infinity \cite{Chandrasekhar:1992}. Due to spherical symmetry, these orbits are planer and generate a photon sphere around the black hole. Photons, coming from the far distance source, approach the black hole with some impact parameter and get deflected symmetrically to infinity, meanwhile reaching a minimum distance near the black hole. The impact parameter $u$ is related to the closest approach distance $x_0$, which follows from intersecting the effective potential $V_{\text{eff}}(x)$ with the energy of photon $\mathcal{E}^2$, this reads as
\begin{equation}
V_{\text{eff}}(x)=\mathcal{E}^2\quad \Rightarrow \quad u\equiv \frac{\cal L}{\cal E} =\sqrt{\frac{C(x_0)}{A(x_0)}}.
\end{equation} 
For $x_0=x_m$, the corresponding impact parameter is $u_m$, which satisfy Eq.~(\ref{co}). The gravitational deflection angle of light is described as the angle between the asymptotic incoming and outgoing trajectories, and reads as \cite{weinberg:1972,Virbhadra:1999nm}
\begin{eqnarray}
\alpha_D (x_0) &=& I(x_0)-\pi,\label{def}
\end{eqnarray}
where 
\begin{equation}\label{def3}
I(x_0)=2\int_{x_0}^\infty \frac{d\phi}{dx}dx,\qquad
 \frac{d\phi}{dx}=\frac{1}{\sqrt{A(x)C(x)}\sqrt{\frac{C(x)A(x_0)}{C(x_0)A(x)}-1}}.
\end{equation}
It is worthwhile to note that the impact parameter coincides with the distance of closest approach only in the limit of vanishing deflection angle. Due to spacetime symmetries, the total change in $\phi$ as $x$ decreases from $\infty$ to its minimum value $x_0$ and then increases again to $\infty$ is just twice its change from $\infty$ to $x_0$. In the absence of a black hole, photons will follow the straight line trajectory and this change in $\phi$ is simply $\pi$ and therefore by Eq.~(\ref{def}) the deflection angle is identically zero.
The deflection angle $\alpha_D (x_0)$ monotonically increases as the distance of minimum approach $x_0$ decreases, and becomes higher than $2\pi$ resulting in the complete loops of the light ray around the
black hole  before escaping to the observer for $x_0\simeq x_m$  (or  $ u\simeq u_m$) and leads to the set of infinite or relativistic source's images \cite{weinberg:1972,Virbhadra:1999nm}. Only for the $x_0=x_m$ (or $u=u_m$), deflection angle diverges logarithmically, whereas, photons with impact parameters $u<u_m$ get captured by the black hole and fall into the horizon. We consider the sign convention such that for $\alpha_D (x_0)>0$, light bend toward the black hole, whereas for $\alpha_D(x_0)<0$, light bend away from it. We are interested in the deflection of light in the strong-field limit, viz., light rays passing close to the photon sphere. In the strong deflection limit, we can expand the deflection angle near the photon sphere, where it diverges \cite{Bozza:2002zj}. For this purpose, we define a new variable \cite{Bozza:2002zj} 
\begin{eqnarray}
z &=& \frac{A(x)-A(x_0)}{1-A(x_0)},
\end{eqnarray}
the integral (\ref{def}) can be re-written as 
\begin{eqnarray}
I(x_0) &=& \int_{0}^{1}R(z,x_0) f(z,x_0) dz,\label{def1}
\end{eqnarray}
with 
\begin{eqnarray}
R(z,x_0) &=&  \frac{2\sqrt{C(x_0)}(1-A(x_0))}{C(x)A'(x)},\\
f(z,x_0) &=& \frac{1}{\sqrt{A(x_0)-\frac{A(x)}{C(x)}C(x_0)}}\label{f(x)},
\end{eqnarray}
where functions without subscript ``$0$" are evaluated at $x=A^{-1}[(1-A(x_0))z+A(x_0)]$ \cite{Bozza:2002zj}. Equation~(\ref{f(x)}), on performing the Taylor series expansion of the function within square root, reduces to 
\begin{eqnarray}
f_0(z,x_0) &=& \frac{1}{\sqrt{\zeta(x_0) z + \beta(x_0) z^2}},
\end{eqnarray}
where 
\begin{eqnarray}
\zeta(x_0) &=& \frac{1-A(x_0)}{A'(x_0) C(x_0)}\left[\left(C^\prime(x_0) A(x_0)-A^\prime(x_0) C(x_0)\right)\right],\\
\beta(x_0) &=& \frac{\left(1-A(x_0)\right)^2}{2 A^\prime(x_0){^3} C(x_0)^2}\Big[2C(x_0)C^\prime(x_0) A^\prime(x_0){^2} +  A(x_0)A^\prime(x_0) C(x_0)C^{\prime\prime}(x_0)   \nonumber\\
&&-C(x_0)C^\prime(x_0) A(x_0) A^{\prime\prime}(x_0) -2C^\prime(x_0){^2}A(x_0)A^\prime(x_0) \Big].
\end{eqnarray}
And 
\begin{eqnarray}
R(z,x_0) &=& \frac{P+Qz}{x_0^3c1^2 (-4 \pi \tilde{\alpha} + 
    x_0^3 (-1 + c1))^4}
\end{eqnarray}
with
\begin{eqnarray}
P &=& 2 x_0^3 c1 (-1 +c1) \Big[2(-1 + c1)x_0^{12}+\pi \tilde{\alpha}(-68+52c1) x_0^9 +\pi^2  \tilde{\alpha}^2(-696 + 344c1 )x_0^6 \nonumber\\
&&+ \pi^3\tilde{\alpha}^3(- 1952+384 c1)x_0^3 
-512 \pi^4 \tilde{\alpha}^4 \Big],\\
Q &=& 2 x_0^3 c1 (-1 +c1) \Big[ 
   3(-1 +c1)x_0^{12} +
   2 \pi  \tilde{\alpha} (-63 + 
      51c1)x_0^9  +
   8 \pi^2 \tilde{\alpha}^2 (-201 + 
      111c1)x_0^6 \nonumber\\
&&+ 32 \pi^3\tilde{\alpha}^3 (-183 + 39c1)x_0^3-1536 \pi^4\tilde{\alpha}^4 \Big],\\
c1 &=& \sqrt{1 + \frac{16 \pi\tilde{\alpha}}{x_m^3}}.
\end{eqnarray}

\begin{figure}[t]
	\begin{center}
		\begin{tabular}{c c}	
			\includegraphics[scale=0.75]{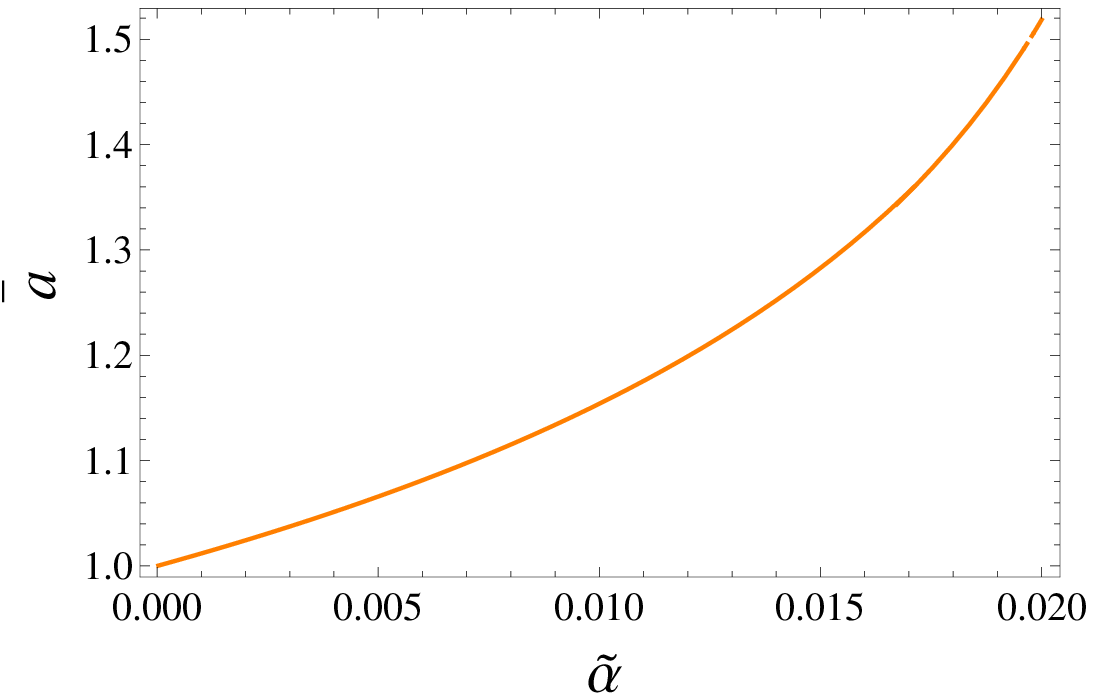}&
			\includegraphics[scale=0.75]{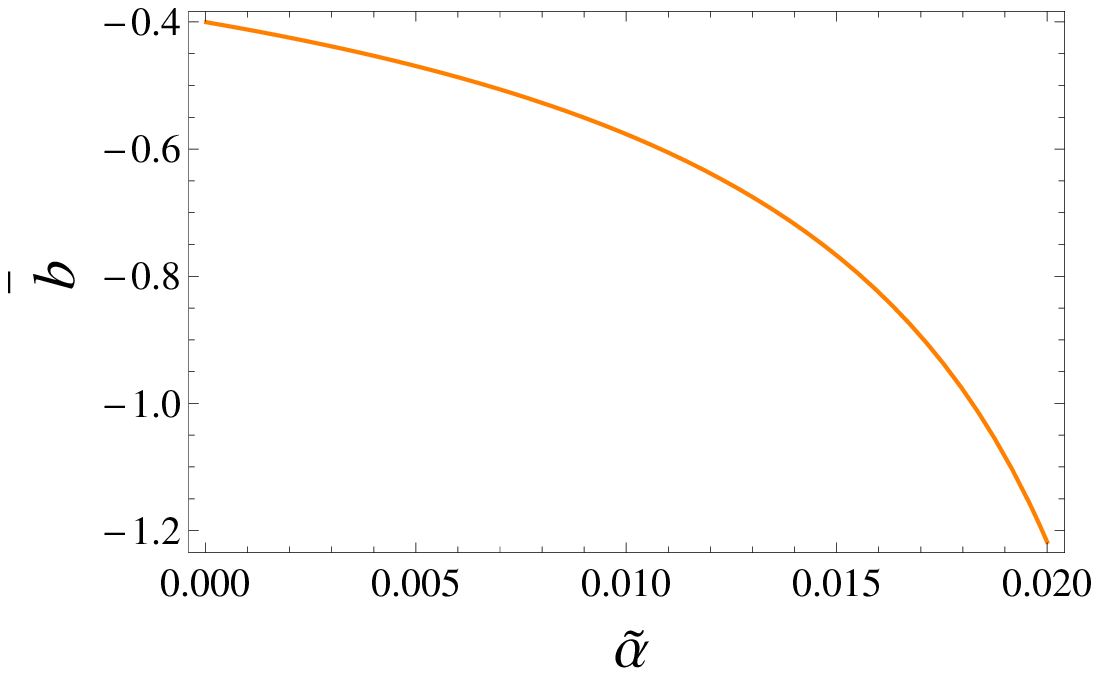}
		\end{tabular}
		\caption{Plot showing the strong lensing coefficients $\bar{a}$ and $\bar{b}$ as function of $\tilde{\alpha}$.}\label{plot2}
	\end{center}
\end{figure}
\begin{center}
\begin{figure}
\begin{tabular}{c c}
     \includegraphics[scale=0.75]{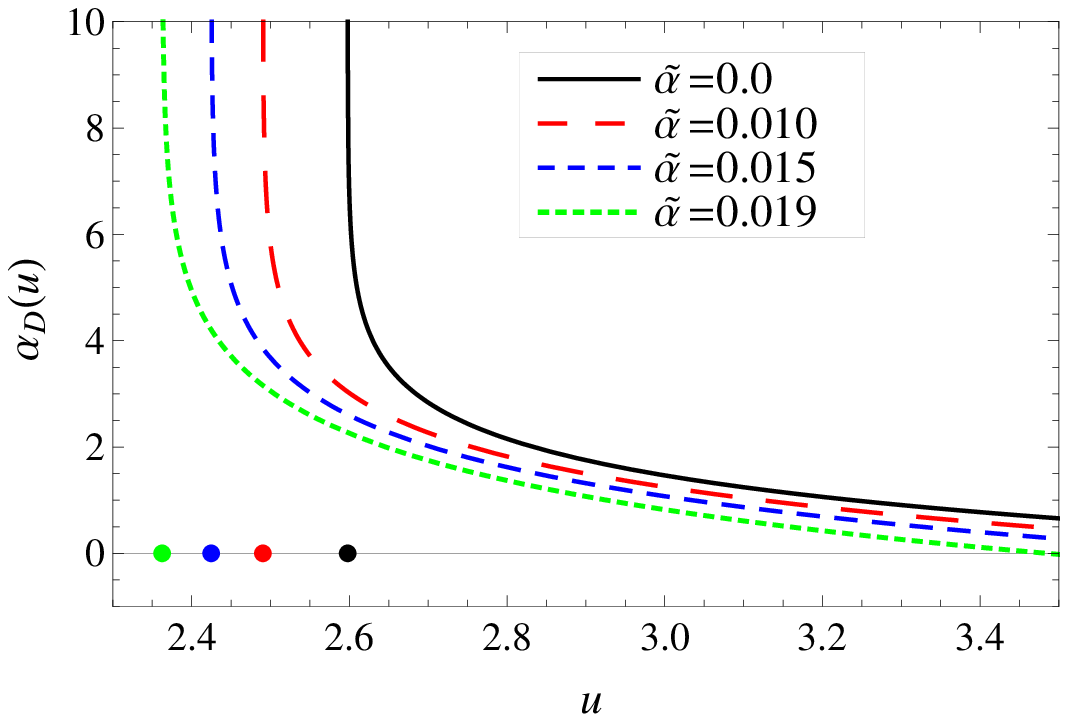}&
     \includegraphics[scale=0.75]{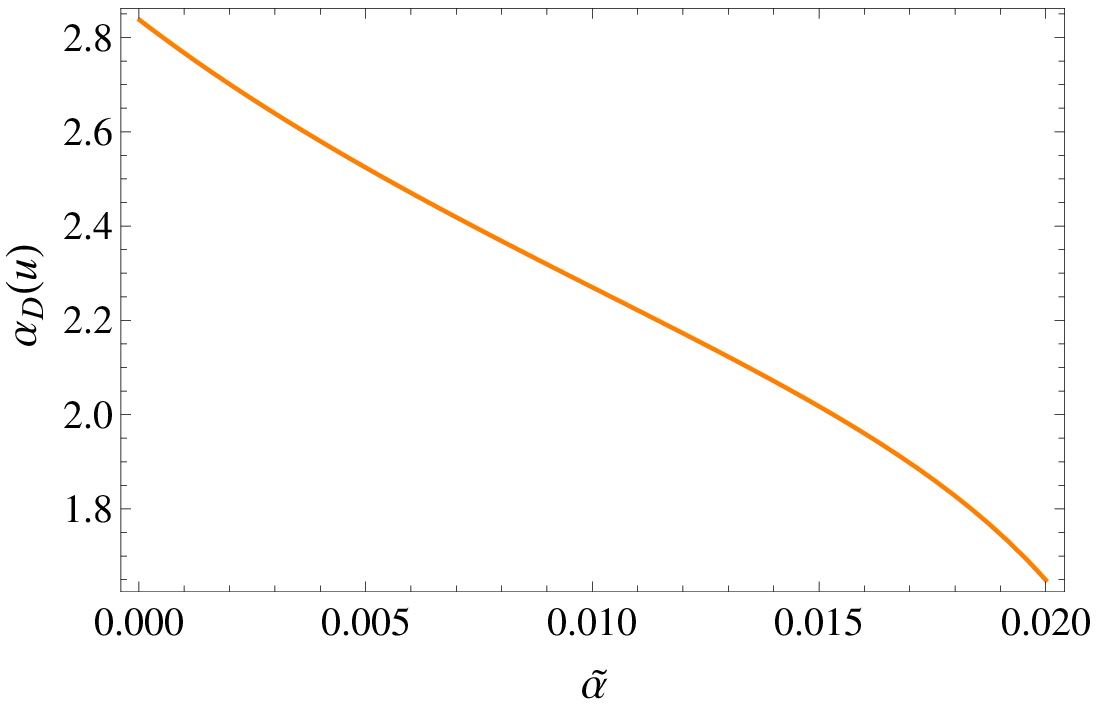}
     \end{tabular}	     
	\caption{Left Panel: plot showing the behavior of deflection angle $\alpha_D(u)$ for strong-gravitational lensing with impact parameter $u$ for different values of $\tilde{\alpha}$. The colored points on the horizontal axis correspond to the impact parameter $u=u_m$, for which deflection angle diverges. Right Panel:  deflection angle $\alpha_D(u)$ variation with $\tilde{\alpha}$ for $u=2.7$.}\label{plot3}
\end{figure}
\end{center}

In Eq.~(\ref{def1}), $R(z,x_0)$ is regular for all values of $z$ and $x_0$, however, $f(z,x_0)$ diverges for $x=x_0$ or $z\to 0$. With the above given definitions, the integral in Eq.~(\ref{def}) can be split into two, diverging and regular, parts as follows \cite{Bozza:2002zj}
\begin{eqnarray}
I(x_0) &=& I_D(x_0)+I_R (x_0),
\end{eqnarray}
such that 
\begin{eqnarray}
I_D(x_0) &=& \int_0 ^1 R(0,x_m)f_0(z,x_0)dz,\label{divergent}\\
I_R(x_0) &=& \int_0 ^1 \Big(R(z,x_0)f(z,x_0)-R(0,x_m)f_0(z,x_0)\Big)dz.\label{regular}
\end{eqnarray}
The integral $I_D(x_0)$ converges for $x_0 \ne x_m$, as $f_0(x_0)=1/\sqrt{z}$. But when $x_0=x_m$ the integral $I_D(x_0)$ has logarithmic divergences as $\phi(x_0)=0 $ and $f_0(x_0)=1/z$. Integral $I_R(x_0)$ is the regularized term with the divergence subtracted.
Solving the integrals in Eqs.~(\ref{divergent}) and (\ref{regular}), the deflection angle can be simplified  \cite{Bozza:2002zj} to 
\begin{eqnarray}\label{defangle}
\alpha_D(u) &=& -\bar{a} \log\Big(\frac{u}{u_m}-1\Big)+\bar{b} + O(u-u_m),
\end{eqnarray} 
where
\begin{align}
\bar{a}&=\frac{R(0,x_m)}{2\beta(x_m)}=\frac{8c1^4\left( 8 \pi \tilde{\alpha} + x_m^2(1-c1) \right)^2}{\left(-c1 + 3\right) \left(-1 + c1\right)^2 \left(x_m^{4} c1 + 4 \pi\tilde{\alpha} \left(-9 + 4 x_m c1 \right)\right)},\\
\bar{b}&= -\pi + I_R(x_m) + \bar{a} \log \Big( \frac{2\beta(x_m)}{A(x_m)}\Big),
\end{align}
and
\begin{eqnarray}
\beta(x_m) &=& \frac{\left(-1 + c1\right)^2 \left(x_m^{4} c1 + 
   4 \pi \tilde{\alpha} \left(-9 + 
      4 x_m c1 \right)\right)}{4 c1^3 \left( 8 \pi\tilde{\alpha} + x_m^2(1-c1) \right)^2}.
\end{eqnarray}
In this case $I_R(x_m)$ can not be calculated analytically, so it has been calculated numerically. $\bar{a}$ and $\bar{b}$ are called the strong deflection limit coefficients, which depend on the metric functions evaluated at the $x_m$.  It is evident that $\bar{a}$ increases whereas $\bar{b}$ decreases with the $\tilde{\alpha}$ (cf. Fig.~\ref{plot2}),  and in the limit $\tilde{\alpha}\to 0$, they smoothly retain the values for the Schwarzschild black hole, viz., $\bar{a}=1$ and $\bar{b}=-0.4002$. The deflection angle $\alpha_D(u)$ in the strong deflection limits for the static spherically symmetric EGB black holes (\ref{NR}) are depicted  in Fig.~(\ref{plot3}). The deflection angle $\alpha_D(u)$ diverges for $u=u_m$ and steeply falls with $u$ (cf. Fig.~\ref{plot3}). It is evident that for fixed values of $u$, deflection angle decrease with increasing GB coupling parameter $\tilde{\alpha}$; Schwarzschild black hole cause larger deflection angle than the EGB black hole. It is worthwhile to note that the presented results are valid only in the strong deflection limit $u\gtrsim u_m$, whereas for $u>>u_m$, the strong deflection limit is not a valid approximation. There exist a value of $x_0=x_z$ or $u=u_z$ for which deflection angle becomes zero.

\section{Lensing observables}\label{sec4}
Once we have known the deflection angle due to strong gravitational lensing Eq.~(\ref{defangle}), we can easily calculate the image positions using the lens equation. Lens equation, establishing a relation between the  observational setup geometry, namely the positions of source $S$, observer $O$ and the black hole $L$ in a given coordinate system, and the position of the lensed images in the observer's sky, is given by \cite{Bozza:2001xd}
\begin{eqnarray}
D_{OS}\tan\beta &=& \frac{D_{OL}\sin\theta - D_{LS}\sin(\alpha-\theta)}{\cos(\alpha-\theta)},\label{lensEq}
\end{eqnarray}
where $\beta$ and $\theta$ are the angular separations, respectively, of the source and the image from the black hole. The distance between the source and black hole is $D_{LS}$, whereas distance from the observer to the source and black hole is $D_{OS}$ and $D_{OL}$ respectively; all distances are expressed in terms of the Schwarzschild radius $x_s=R_s/2M$. Photon emitted with a impact parameter $u$ from the source approaches the black hole and is received by an observer. The deflection angle $\alpha_D$ is identified as the angle between the tangents to the emission and detection directions. Since it diverges for $u=u_m$ and the light rays perform several loops around the black hole before escaping to the observer, therefore, we can replace $\alpha_D$ by $2n\pi + \Delta \alpha_n $, where $n$ is the positive integer number corresponding to the winding number of loops around black hole and $\Delta \alpha_n $ is the offset of the deflection angle. For the case of a far distant observer and source, and their nearly perfect alignment with the black hole, the lens equation (\ref{lensEq}) can be simplified as \cite{Bozza:2001xd,Bozza:2008ev}
\begin{eqnarray}\label{lenseq}
\beta &=& \theta -\frac{D_{LS}}{D_{OS}}\Delta \alpha_n.
\end{eqnarray}
However, the lens equation has also been defined in more general setup \cite{Frittelli:1998hr,Perlick:2004zh,Perlick:2003vg,Eiroa:2005vd,Li:2015vqa,Frittelli:1999yf,Eiroa:2003jf,Eiroa:2004gh,Bozza:2007gt}. One can notice that in Eq.~(\ref{lenseq}), only the offset angle $\Delta \alpha_n $ comes into the lens equation rather than the complete deflection angle. To get the offset deflection angle for $n^{th}$ relativistic image, $\Delta \alpha_n $, we first solve the $\alpha_D(\theta_n^0)=2n\pi$, where $\theta_n^0$ is the image position for the $\alpha_D=2n\pi$, this yields
\begin{equation}\label{theta0}
\theta_n ^0=\frac{u_m}{D_{OL}}(1+e_n),
\end{equation} 
with
\begin{equation}
e_n= e^{{\bar{b}-2n\pi}/{\bar{a}}}.
\end{equation}
Now, making a Taylor series expansion of the deflection angle about $(\theta_n ^0)$ to the first order, gives
\begin{eqnarray}\label{Taylor}
\alpha_D(\theta) &=& \alpha_D(\theta_n ^0) +\frac{\partial \alpha_D(\theta)}{\partial \theta } \Bigg |_{\theta_n ^0}(\theta-\theta_n ^0)+O(\theta-\theta_n ^0),
\end{eqnarray}
using $\Delta \theta_n=\theta-\theta_n ^0 $ and the deflection angle Eq.~(\ref{defangle}), the Eq.~(\ref{Taylor}) becomes 
\begin{eqnarray}
\Delta\alpha_n &=& -\frac{\bar{a}D_{OL}}{u_m e_n}\Delta\theta_n.
\end{eqnarray}
Neglecting the higher-order terms, the lens equation finally gives the position of $n^{th}$ image \cite{Bozza:2002zj}
\begin{equation}
\theta_n\simeq \theta_n ^0 + \frac{D_{OS}}{D_{LS}}\frac{u_me_n}{D_{OL}\bar{a}}(\beta-\theta_n ^0),\label{defcorrection}
\end{equation}
for $\beta=\theta_n ^0$, viz., the image position coincides with the source position, the correction to the $n^{th}$ image position identically vanishes. Though Eq.~(\ref{defcorrection}) gives source's images on the same side of source ($\theta >0$), we can replace $\beta$ by $-\beta$ in order to get the images on the other side. The brightness of the source's images will be magnified due to the gravitational lensing, the magnification of $n^{th}$ loop image is defined as \cite{Bozza:2002zj}
\begin{eqnarray}
\mu_n &=& \left(\frac{\beta}{\theta} \;  \;\frac{d\beta}{d\theta} \Bigg|_{\theta_n ^0}\right)^{-1}.\label{magn}
\end{eqnarray}
The brightness magnification decreases steeply with the order $n$ of the images, such that unless the source is almost perfectly aligned with the black hole and the observer, these images will be very faint as a result of high demagnification. Also Eq.~(\ref{magn}) infers that for the perfect alignment of the source and the black hole, $\beta\to 0$, the brightness magnification diverges. 
Considering only the outermost one-loop image $\theta_1$ as distinguishable as a single image from the remaining inner packed images $\theta_{\infty}$, we can have three characteristic observables \cite{Bozza:2002zj}
\begin{eqnarray}
\theta_\infty &=& \frac{u_m}{D_{OL}},\\
s &=& \theta _1-\theta _\infty = \theta_\infty \;\; e^{\frac{\bar{b}-2\pi}{\bar{a}}},\label{s}\\
r &=&  \frac{\mu_1}{\sum{_{n=2}^\infty}\mu_n } = e^{\frac{2 \pi}{\bar{a}}},\qquad r_{\text{mag}}= 2.5 \log(r) = \frac{5\pi}{\bar{a}\ln 10}.\label{obs}
\end{eqnarray}
where 
\begin{eqnarray}
u_m &=& \frac{2\sqrt{2\pi\tilde{\alpha}}}{\sqrt{1+8\pi\tilde{\alpha}-c1}}.
\end{eqnarray}
Here, $\theta_{\infty}$ is the angular radius of the photon sphere, i.e., the position of the innermost packed images, $s$ is the angular separation between the one-loop image and the images at $\theta_\infty$, and $r$ is the ratio of the brightness flux from the outermost relativistic image at $\theta_1$ to those from the remaining
relativistic images at $\theta_{\infty}$. It must be noted that, in contrary to $\theta_{\infty}$ and $s$, $r_{\text{mag}}$ is independent of the black hole distance from the observer.

Now, we consider the supermassive black hole candidates at the galactic center of the Milky Way and the nearby galaxy Messier 87, respectively, Sgr A* and M87*, as the static spherically symmetric EGB black hole described by metric (\ref{NR}).  
\begin{figure}
	\begin{center}	
		\begin{tabular}{c c}
			\includegraphics[scale=0.84]{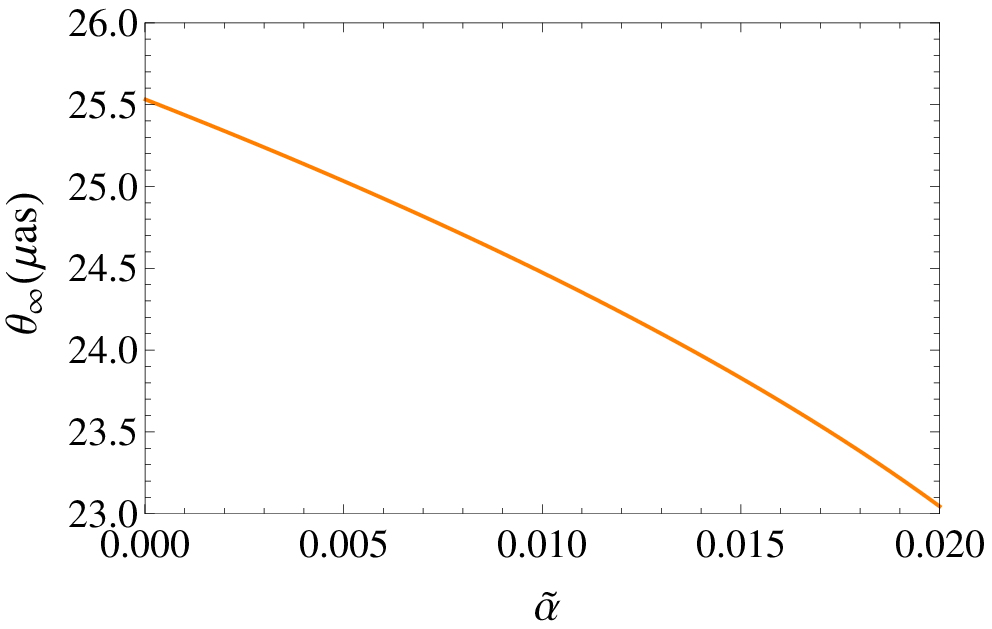}&
			\includegraphics[scale=0.84]{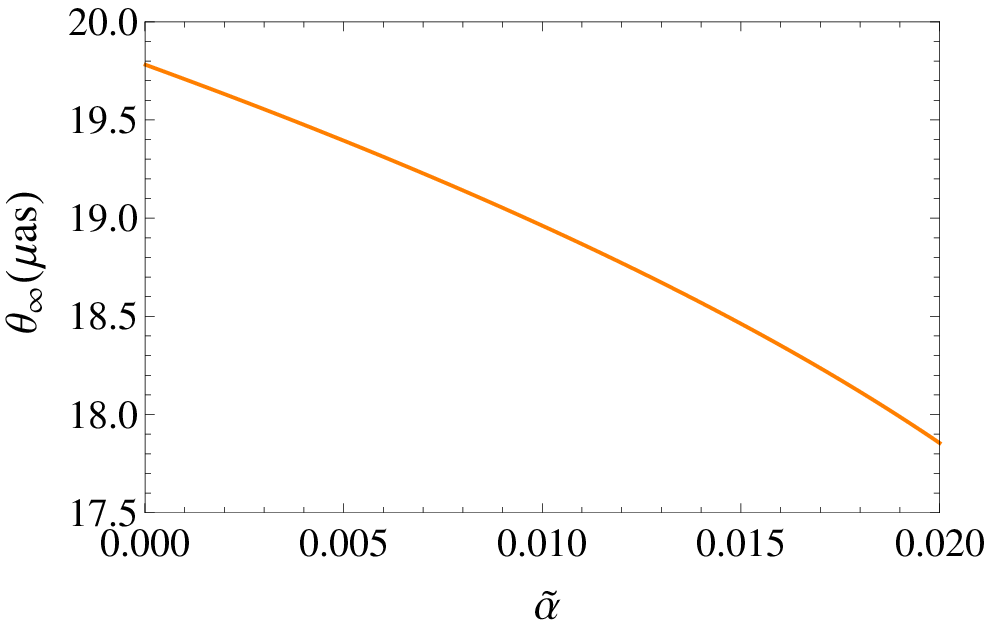}\\
			\includegraphics[scale=0.84]{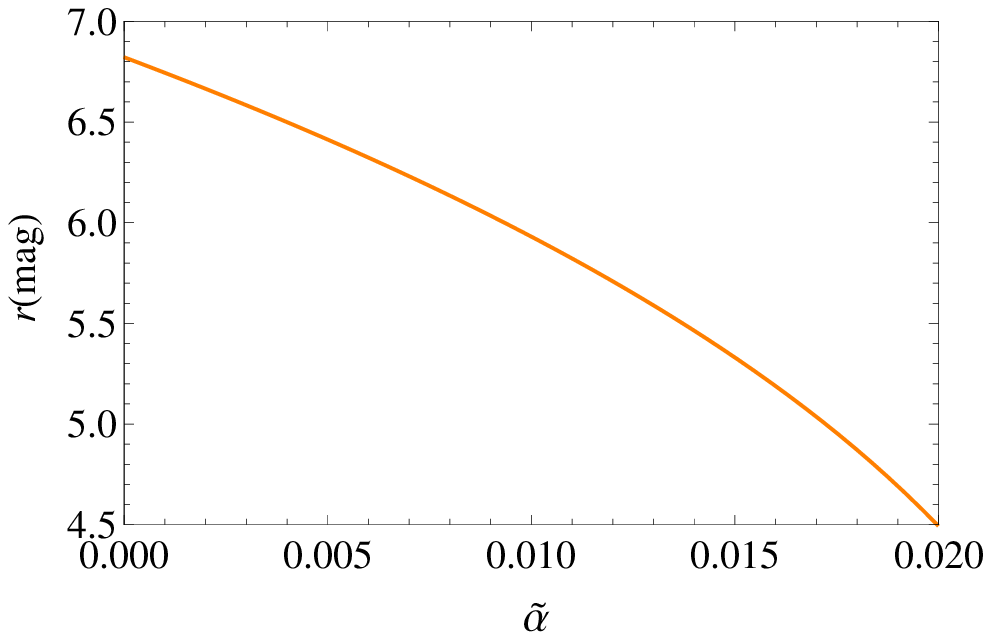}&
	    	\includegraphics[scale=0.84]{magr.eps}\\	
			\includegraphics[scale=0.84]{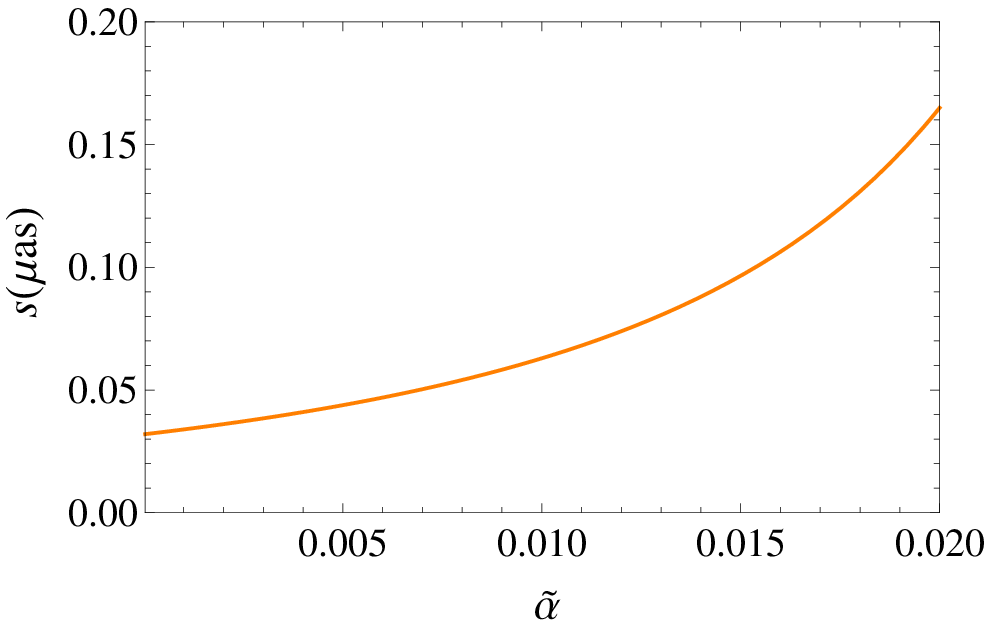}&
			\includegraphics[scale=0.84]{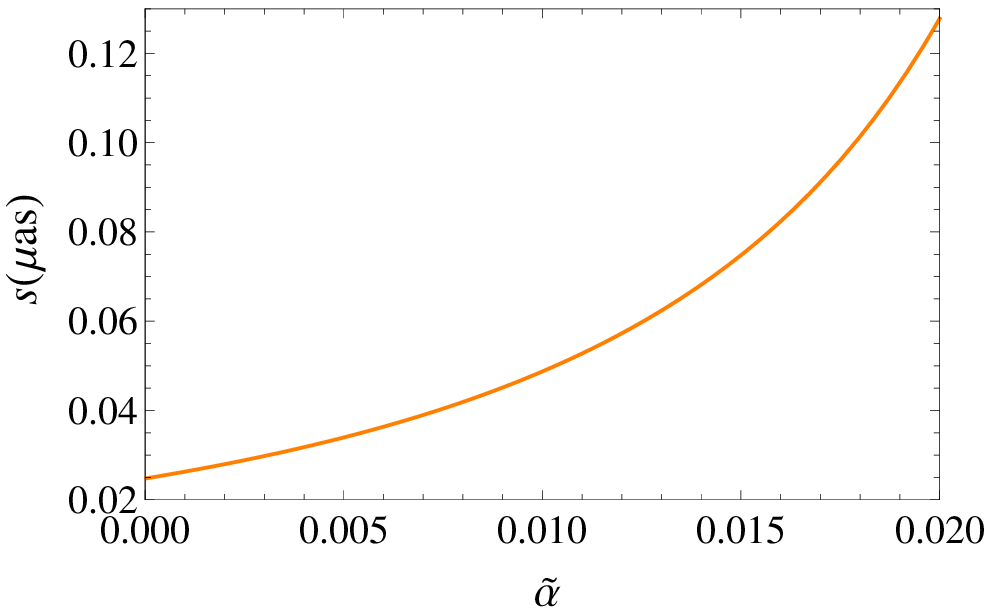}	
		\end{tabular}
		\caption{Plot showing the behavior of lensing observables,  $\theta_\infty$, $r$, and $s$ as a function of $\tilde{\alpha}$ for Sgr A* (left panel) and M87* (right panel) black holes. }\label{plot4}
	\end{center}	
\end{figure} 
Based on the latest observational data, we have taken their masses and distances from the Earth as, $M=3.98\times 10^6 M_{\odot}$ and $D_{OL}=7.97$ kpc for Sgr A* \cite{Do:2019txf}, and $M=(6.5\pm 0.7)\times 10^9 M_{\odot}$ and $D_{OL}=(16.8\pm 0.8)$ Mpc for M87* \cite{Akiyama:2019cqa}. 
It is interesting to estimate the correction due to the GB coupling parameter $\tilde{\alpha}$ by comparing the results of EGB black hole with those  for the Schwarzschild black hole (cf. Table~\ref{table3}). 
The lensing observables for the Sgr A* and M87* black holes are shown in Fig.~(\ref{plot4}) as a function of $\tilde{\alpha}$. Figure~(\ref{plot4}) and Table~\ref{table3} clearly infer that the relativistic images have largest angular separation for small values of $\tilde{\alpha}$. By measuring the lensing observables, namely $\theta_\infty$, $s$, and $r$, for the EGB black hole and inverting the Eqs.~\ref{s} and (\ref{obs}), we can calculate the strong deflection coefficients $\bar{a}$ and $\bar{b}$. Then the theoretically predicted values can be compared with the values inferred from the observational data to find the black hole parameters.    

\begin{table}
	\centering
	\begin{tabular}{p{1cm} p{1.1cm} p{1.1cm} p{1.5cm} p{1.1cm} p{1.1cm}p{1.5cm} p{1.2cm}p{1.2cm} p{1.2cm}}
		\hline\hline
\multicolumn{1}{c}{} & \multicolumn{3}{c}{Sgr A*} & \multicolumn{3}{c}{M87*}  & \multicolumn{3}{c}{Lensing Coefficients}\\
		{$\tilde{\alpha}$ } & {$\theta_\infty $ ($\mu$as)} & {$s$ ($\mu$as) } &  {$r_m$}& {$\theta_\infty $ ($\mu$as)} & {$s$ ($\mu$as) } &  {$r_m$}&{$\bar{a}$}&{$\bar{b}$}&{$u_m/R_s$}\\\hline
		
		$0.0$ &  $25.530$  & $0.031$  & $6.825$&  $19.780$  & $0.024$  & $6.825$ & $1.0$ & $-0.401$ & $2.597$\\ 
		
		$0.005$     &  $25.032$     &  $0.043$  & $6.412$&  $19.395$  & $0.034$  & $6.412$ & $1.063$ &$-0.469$ & $2.547$\\ 
		
		$0.01$       & $24.473$   &  $0.063$  & $5.931$ &  $18.961$  & $0.048$  & $5.931$ & $1.150$ & $-0.576$ &$2.490$\\
		
		$0.015$        & $23.829$  &  $0.096$  & $5.330$&  $18.462$  & $0.074$  & $5.330$ & $1.279$ &$-0.767$ & $2.424$\\
		
		$0.019$       & $23.218$  &  $0.146$  & $4.691$ &  $17.988$  & $0.113$  & $4.691$ & $1.454$ & $-1.083$ &$2.362$\\
		\hline\hline
	\end{tabular}
	\caption{Strong-lensing observables for the black hole Sgr A*  and M87*, and lensing coefficients for various values of $\tilde{\alpha}$. } 
	\label{table3}  
\end{table}

\section{Weak gravitational lensing}\label{sec5}
Gibbon and Werner \cite{Gibbons:2008rj} invoked the Gauss-Bonnet theorem, in the context of optical geometry to calculate the deflection angle of light in the weak-field limits of the spherically symmetric black hole spacetime \cite{Carmo}. Later, considering the source and observers at finite distances from the black hole, corrections in the deflection angle due to static spherically symmetric black hole were calculated by  Ishihara \textit{et al.} \cite{Ishihara:2016vdc,Ishihara:2017}, which is further generalized by Ono \textit{et al.} \cite{Ono:2017pie} for the stationary and axisymmetric black holes. Since then, their method have been extensively used for varieties of black hole spacetimes \cite{Ovgun:2019wej,Crisnejo1:2018uyn,Ovgun:2018fnk,Ovgun:2018tua,Javed:2019rrg,Javed:2019ynm,Javed:2019kon,Crisnejo:2019xtp,Crisnejo:2018ppm,Zhu:2019ura,Kumar:2019pjp}. We follow their approach to calculate the light deflection angle in the weak-field limit caused by the static and spherically symmetric EGB black hole. 

\begin{figure}
\centering
	\includegraphics[scale=0.69]{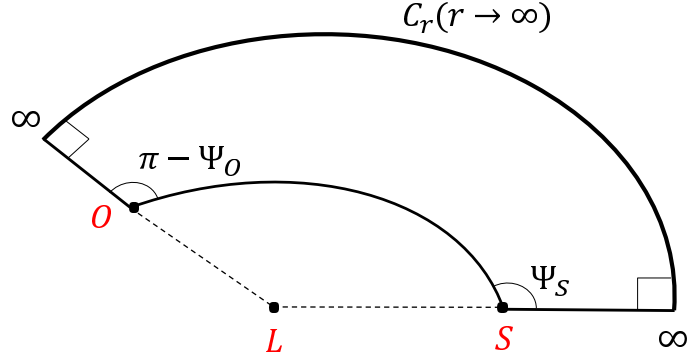}
	\caption{Schematic figure for the quadrilateral ${}_O^{\infty}\Box_{S}^{\infty}$ embedded in the curved space. }\label{lensing1}
\end{figure}

\begin{figure}
	\centering
		\includegraphics[scale=0.9]{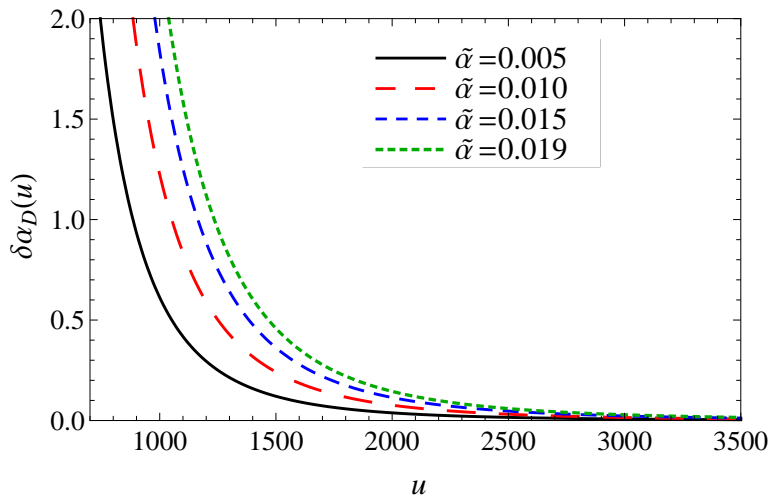}
	\caption{The corrections in the deflection angle $\delta\alpha_D =\left.\alpha_D\right|_{\text{Schw}}-\alpha_D$ for weak-gravitational lensing around EGB black hole with $b_s=10^{-4}$, $b_o=0$ and varying $u$;  $\delta\alpha_D$ is in units of $\mu$as.}\label{Defweak}
\end{figure}

Considering a coordinate system centered at the black hole ($L$), and assuming the observer ($O$) and the source ($S$) at the finite distances from the black hole 
(cf. Fig.~\ref{lensing1}), we can define the deflection angle at the equatorial plane as follow \cite{Ishihara:2016vdc,Ono:2017pie}
\begin{equation}
\alpha_D=\Psi_O-\Psi_S+\Phi_{OS},
\end{equation}
where, $\Phi_{OS}$ is the angular separation between the observer and the source, $\Psi_S$ and $\Psi_O$, respectively, are the angle made by light rays at the source and observer. The quadrilateral ${}_O^{\infty}\Box_{S}^{\infty}$, which is made up of spatial light ray curves from source to the observer, a circular arc segment $C_r$ of coordinate radius $r_C$ $(r_C\to\infty)$, and two outgoing radial lines from $O$ and $S$, is embedded in the  3-dimensional Riemannian manifold $^{(3)}\mathcal{M}$ defined by optical metric $\gamma_{ij}$ (cf. Fig.~\ref{lensing1}).
The surface integral of the Gaussian curvature of the two-dimensional surface of light propagation in this manifold gives the light deflection angle \cite{Ishihara:2016vdc,Ono:2017pie}
\begin{equation}
\alpha_D=-\int\int_{{}_O^{\infty}\Box_{S}^{\infty}} K dS.\label{deflectionangle}
\end{equation}
Solving Eq.~(\ref{NR}) for the null geodesics $ds^2=0$, we get
\begin{equation}
dt= \pm\sqrt{\gamma_{ij}dx^i dx^j},
\end{equation}
with
\begin{equation}
\gamma_{ij}dx^i dx^j=\frac{1}{f(r)^2}dr^2+\frac{r^2}{f(r)}\left(d\theta^2+\sin^2\theta\, d\phi^2\right). \label{metric3}
\end{equation}
Gaussian curvature of the surface of light propagation is defined as \cite{Werner:2012rc}
\begin{eqnarray}
K&=&\frac{{}^{3}R_{r\phi r\phi}}{\gamma},\nonumber\\
&=&\frac{1}{\sqrt{\gamma}}\left(\frac{\partial}{\partial \phi}\left(\frac{\sqrt{\gamma}}{\gamma_{rr}}{}^{(3)}\Gamma^{\phi}_{rr}\right) - \frac{\partial}{\partial r}\left(\frac{\sqrt{\gamma}}{\gamma_{rr}}{}^{(3)}\Gamma^{\phi}_{r\phi}\right)\right),\label{gaussian}
\end{eqnarray}
where $\gamma=\det(\gamma_{ij})$. For EGB black hole metric (\ref{NR}), in the weak-field limits,  Eq.~(\ref{gaussian}) simplifies to
\begin{eqnarray}
K&=&-\frac{2M}{r^3}+\frac{3M^2}{r^4}+\frac{640M^2\pi\alpha}{r^6}-\frac{1152M^3\pi\alpha}{r^7}+\mathcal{O}\left(\frac{M^3\alpha^2}{r^9},\frac{M^4\alpha^2}{r^{10}}\right).
\end{eqnarray}
The surface integral of Gaussian curvature over the closed quadrilateral ${}_O^{\infty}\Box_{S}^{\infty}$ reads \cite{Ono:2017pie}
\begin{equation}
\int\int_{{}_O^{\infty}\Box_{S}^{\infty}} K dS= \int_{\phi_S}^{\phi_O}\int_{\infty}^{r_0} K \sqrt{\gamma}dr d\phi,\label{Gaussian}
\end{equation}
where $r_0$ is the distance of closest approach to the black hole. In the weak-field approximation, the light orbits equation can be considered as   \cite{Crisnejo:2019ril}
\begin{equation}
b=\frac{\sin\phi}{u} + \frac{M(1-\cos\phi)^2}{u^2}-\frac{M^2(60\phi\,\cos\phi+3\sin3\phi-5\sin\phi)}{16u^3}+\mathcal{O}\left( \frac{M^2\alpha}{u^5}\right)
,\label{uorbit}
\end{equation} 
where $b=1/r$, and $u$ is the impact parameter. The integral Eq.~(\ref{Gaussian}) can be recast as
\begin{equation}
\int\int_{{}_O^{\infty}\Box_{S}^{\infty}} K dS= \int_{\phi_S}^{\phi_O}\int_{0}^{b}-\frac{K\sqrt{\gamma}}{b^2}db d\phi,
\end{equation}
which for the metric Eq.~(\ref{metric3}) reads as
	\begin{eqnarray}
	\int\int K dS&=&\left(\cos^{-1}ub_o+\cos^{-1}ub_s\right)\Big(\frac{15M^2}{4u^2}-\frac{60M^2\pi\alpha}{u^4}\Big)\nonumber\\
	&+&\left(\sqrt{1-u^2b_o^2}+\sqrt{1-u^2b_s^2}\right)\Big(\frac{2M}{u}+\frac{128M^3}{6u^3}-\frac{14852 M^3\pi\alpha }{25u^5}\Big) \nonumber\\
	&+& \left(b_o\sqrt{1-u^2b_o^2}+b_s\sqrt{1-u^2b_s^2} \right)\Big(-\frac{M^2}{4u} -\frac{60M^2\pi\alpha }{u^3}\Big)\nonumber\\
	&+& \left(b_o^2\sqrt{1-u^2b_o^2} +b_s^2\sqrt{1-u^2b_s^2}  \right)\Big(\frac{ M^3}{6u}-\frac{7426 M^3\pi\alpha}{25u^3}\Big)\nonumber\\
	&+&\mathcal{O}\left(\frac{M^4}{u^4},\frac{M^4\alpha}{u^6}\right),\label{Gaussian1}
	\end{eqnarray}
where, $b_o$ and $b_s$, respectively, are the reciprocal of  the distances of observer and source from the black hole, i.e., $b_o=1/r_o$ and $b_s=1/r_s$. We have used the approximation $\cos\phi_o=-\sqrt{1-u^2b_o^2},\;\; \cos\phi_s=\sqrt{1-u^2b_s^2}$, the relative negative sign is because the source and the observer are at the opposite sides to the black hole. In the limits of far distant observer and source, $b_o\to 0$ and $b_s\to 0$, the deflection angle for the EGB black hole in the weak-field limits reads as follows
\begin{eqnarray}
\alpha_D&=&\frac{4M}{u}+\frac{15\pi M^2}{4u^2}-\frac{60M^2\pi^2\alpha}{u^4}+\frac{128M^3}{3u^3}-\frac{29704 M^3\pi\alpha}{25u^5}+\mathcal{O}\left(\frac{M^4}{u^4},\frac{M^4\alpha}{u^6}\right).\label{defAngle}
\end{eqnarray}
Further, in the limiting case of $\alpha=0$, it reduces as
\begin{equation}
\left.\alpha_D\right|_{\text{Schw}}=\frac{4M}{u}+\frac{15\pi M^2}{4u^2}+\frac{128M^3}{3u^3}+\mathcal{O}\left(\frac{M^4}{u^4} \right),
\end{equation}
which corresponds to the value for the Schwarzschild black hole  \cite{Virbhadra:1999nm,Crisnejo:2019ril}. In terms of the normalized impact parameter and GB coupling parameter, $u\to u/M$ and $\alpha\to\tilde{\alpha}=\alpha/M^2$, the deflection  angle Eq.~(\ref{defAngle}), reads as
\begin{eqnarray}
\alpha_D&=&\frac{4}{u}+\frac{15\pi }{4u^2}-\frac{60\pi^2\tilde{\alpha}}{u^4}+\frac{128}{3u^3}-\frac{29704\pi\tilde{\alpha}}{25u^5}+\mathcal{O}\left(\frac{1}{u^4},\frac{\tilde{\alpha}}{u^6}\right).\label{defAngle1}
\end{eqnarray}
It is clear from Eq.~(\ref{defAngle1}), that the GB coupling parameter $\tilde{\alpha}$ reduces the deflection angle, i.e., in the weak-field limits the EGB black hole leads to smaller deflection angle than the Schwarzschild black hole. We presented the correction in the deflection angle $\delta \alpha_D=\alpha_D-\left.\alpha_D\right|_{\text{Schw}}$ due to $\tilde{\alpha}$ in Table~\ref{T2} and Fig.~\ref{Defweak} for various values of impact parameter $u$. Table~\ref{T2} infers that, for fixed value of impact parameter $u$, the correction $\delta\alpha_D$ increases with the $\tilde{\alpha}$ and is of the order of micro-arc-second. However, for a fixed value of $\tilde{\alpha}$, $\delta\alpha_D$ decreases with $u$.

\begin{table}
	\centering
		\begin{tabular}{|c||c|c|c|c|c|}
		\hline	
		$ $  & $\tilde{\alpha}=0.001$    & $\tilde{\alpha}=0.003$     & $\tilde{\alpha}=0.005$     &$ \tilde{\alpha}=0.01 $   &$ \tilde{\alpha}=0.019 $   \\
		\hline\hline
		$u=1*10^3$&	0.122428& 0.367283& 0.612139& 1.22428& 2.32613 \\  \hline
		$u=2*10^3$&  0.00762716& 0.0228815& 0.0381359& 0.0762718& 0.144916 \\  \hline
		$u=3*10^3$&  0.00150446& 0.0045134& 0.0075223& 0.0150446& 0.0285848 \\  \hline
		$u=4*10^3$& 0.000475089& 0.0014253& 0.00237546& 0.00475093& 0.00902681 \\  \hline		
	\end{tabular}
	\caption{The corrections in the deflection angle $\delta\alpha_D =\left.\alpha_D\right|_{\text{Schw}}-\alpha_D$ for weak-gravitational lensing around the EGB black hole with source at $b_s=10^{-4}$ and observer at $b_o=0$; $\delta\alpha_D$ is in units of $\mu$as. }\label{T2}
\end{table}

\section{Conclusions}\label{sec6}
The GB correction to the Einstein-Hilbert action provides a natural extension to the Einstein's theory of general relativity in $D>4$ dimensional spacetime, however, it is a topological invariant quantity in $D\to 4$, and therefore does not contribute to the gravitational field equations. Recently proposed $4D$ regularized EGB gravity theories, in which  GB term in the gravitational action makes a non-trivial contribution to the field equations in $4D$ and contrary to the  Schwarzschild black hole solution of GR, a black hole in this theory can possess up to two horizons. 

Gravitational lensing is unequivocally a potentially powerful tool for the analysis of strong fields and for testing general relativity. The strong-field limit provides a useful framework for comparing lensing by different  gravities, the $4D$ EGB gravity is a very interesting model to consider  and discuss the observational signatures this quadratic curvature corrected gravity has than the  Schwarzschild black hole of general relativity. In view of this, we investigated the gravitational lensing of light around the $4D$ EGB black hole in the strong deflection limits. Depending on the value of impact parameter $u$, photons get deflected from their straight path and leads to the multiple images of a source, and for the particular value of $u=u_m$, photons follow the circular orbits around the black hole and deflection angle diverges. It is found that the static spherically symmetric EGB black holes lead to smaller deflection angle as compared to the Schwarzschild black hole value, and the deflection angle decreases with the GB coupling parameter $\tilde{\alpha}$. The effect of $\tilde{\alpha}$ on the deflection angle immediately reflect on the relativistic images. Considering the observer and the light source at far distances from the black hole, we calculate the observables for the strong lensing, namely the angular separation between the relativistic source's images and the relative brightness magnitude of the first image. It is noted that, with the increasing GB coupling parameter, the photon circular orbits radii decrease, the angular separation between the set of source images increase, whereas the brightness magnitude decrease. 

We modeled the supermassive black holes Sgr A* and M87,* respectively, in the Galactic Center and at the center of galaxy M87 as the $4D$ EGB black hole (lens), and estimate the lensing observables. The corrections in the angular separation of the images, due to the GB coupling parameter $\alpha$, is of the order of $\mu$as, which is within the limit of current observational outreaches. The weak gravitational lensing around EGB black holes is also discussed and corrections in the deflections angle due to $\tilde{\alpha}$ are calculated, which increase with $\tilde{\alpha}$. 

Investigation of the gravitational lensing around the rotating EGB black holes in the strong and weak field limits is the topic of our future investigation. 

\section{Acknowledgment} S.G.G. would like to thank DST INDO-SA bilateral project DST/INT/South Africa/P-06/2016, SERB-DST for the ASEAN project IMRC/AISTDF/CRD/2018/000042 and also IUCAA, Pune for the hospitality while this work was being done. R.K. would like to thank UGC for providing SRF. S.U.I would like to thank SERB-DST for the ASEAN project IMRC/AISTDF/CRD/2018/000042.

\noindent

\begin{thebibliography}{99}

\bibitem{Gross} 
B.~Zwiebach,
Phys.\ Lett.\  B {\bf 156}, 315 (1985).

\bibitem{Duff:1986pq} 
M.~J.~Duff, B.~E.~W.~Nilsson and C.~N.~Pope,
Phys.\ Lett.\ B {\bf 173}, 69 (1986).

\bibitem{Gross2}
D. J.~Gross and J. H.~Sloan, { Nucl. Phys. B} {\bf 291}, 41 (1987).

\bibitem{Bento}
M.~C.~Bento and O.~Bertolami, {Phys. Lett. B} {\bf 368}, 198 (1996).
\bibitem{Lanczos:1938sf}
C.~Lanczos,
Annals Math.\  {\bf 39} 842 (1938).


\bibitem{Lovelock:1971yv}
D.~Lovelock,
J.\ Math.\ Phys.\  {\bf 12}  498 (1971).


	
\bibitem{Boulware:1985wk} 
D.~G.~Boulware and S.~Deser,
Phys.\ Rev.\ Lett.\  {\bf 55}, 2656 (1985).


\bibitem{Wheeler:1985nh} 
J.~T.~Wheeler,
Nucl.\ Phys.\ B {\bf 268}, 737 (1986).


\bibitem{Tangherlini:1963bw} 
F.~R.~Tangherlini,
Nuovo Cim.\  {\bf 27}, 636 (1963).

	
\bibitem{Glavan:2019inb}
D.~Glavan and C.~Lin,
Phys.\ Rev.\ Lett.\  {\bf 124}, 081301 (2020).


\bibitem{Konoplya:2020juj}
R.~A.~Konoplya and A.~Zhidenko,
Phys. Dark Univ. \textbf{30}, 100697 (2020).

\bibitem{Zhang:2020qew}
Y.~P.~Zhang, S.~W.~Wei and Y.~X.~Liu,
arXiv:2003.10960 [gr-qc].


\bibitem{Hegde:2020xlv}
K.~Hegde, A.~N.~Kumara, C.~L.~A.~Rizwan, A.~K.~M. and M.~S.~Ali,
arXiv:2003.08778 [gr-qc].

\bibitem{Zhang:2020qam}
C.~Y.~Zhang, P.~C.~Li and M.~Guo,
arXiv:2003.13068 [hep-th].

\bibitem{Lu:2020iav}
H.~Lu and Y.~Pang,
Phys. Lett. B \textbf{809}, 135717 (2020).

\bibitem{Churilova}
M.~S.~Churilova,
arXiv:2004.00513 [gr-qc].

\bibitem{Fernandes:2020rpa} 
P.~G.~S.~Fernandes,
Phys. Lett. B \textbf{805}, 135468 (2020).

\bibitem{Konoplya:2020qqh}
R.~A.~Konoplya and A.~Zhidenko,
Phys. Rev. D \textbf{101}, 084038 (2020).

\bibitem{Casalino:2020kbt} 
A.~Casalino, A.~Colleaux, M.~Rinaldi and S.~Vicentini,
arXiv:2003.07068 [gr-qc].

\bibitem{Singh:2020nwo} 
D.~V.~Singh, S.~G.~Ghosh and S.~D.~Maharaj,
arXiv:2003.14136 [gr-qc].

\bibitem{Ghosh1:2020vpc} 
S.~G.~Ghosh and S.~D.~Maharaj,
Phys. Dark Univ. \textbf{30}, 100687 (2020).

\bibitem{Ghosh:2020syx}
S.~G.~Ghosh and R.~Kumar,
arXiv:2003.12291 [gr-qc].


\bibitem{Doneva:2020ped} 
D.~D.~Doneva and S.~S.~Yazadjiev,
arXiv:2003.10284 [gr-qc].

\bibitem{Konoplya:2020ibi} 
R.~A.~Konoplya and A.~Zhidenko,
Phys. Rev. D \textbf{102}, 064004 (2020).

\bibitem{Singh:2020xju} 
D.~V.~Singh and S.~Siwach,
Phys. Lett. B \textbf{808}, 135658 (2020).

\bibitem{Kumar:2020uyz} 
A.~Kumar and R.~Kumar,
arXiv:2003.13104 [gr-qc].

\bibitem{Wei:2020ght} 
S.~W.~Wei and Y.~X.~Liu,
arXiv:2003.07769 [gr-qc].

\bibitem{Kumar:2020owy} 
R.~Kumar and S.~G.~Ghosh,
JCAP \textbf{07}, 053 (2020).

\bibitem{HosseiniMansoori:2020yfj}
S.~A.~Hosseini Mansoori,
arXiv:2003.13382 [gr-qc].


\bibitem{Guo:2020zmf} 
M.~Guo and P.~C.~Li,
Eur. Phys. J. C \textbf{80}, 588 (2020).

\bibitem{Konoplya:2020bxa}
R.~A.~Konoplya and A.~F.~Zinhailo,
arXiv:2003.01188 [gr-qc].


\bibitem{Tomozawa:2011gp}
Y.~Tomozawa,
arXiv:1107.1424 [gr-qc].


\bibitem{Cognola:2013fva} 
G.~Cognola, R.~Myrzakulov, L.~Sebastiani and S.~Zerbini,
Phys.\ Rev.\ D {\bf 88}, 024006 (2013).

\bibitem{Hennigar:2020lsl}
R.~A.~Hennigar, D.~Kubiznak, R.~B.~Mann and C.~Pollack,
JHEP \textbf{07}, 027 (2020).

\bibitem{Mahapatra:2020rds}
S.~Mahapatra,
arXiv:2004.09214 [gr-qc].

\bibitem{Ai:2020peo}
W.~Y.~Ai,
Commun. Theor. Phys. \textbf{72}, 095402 (2020).

\bibitem{Shu:2020cjw}
F.~Shu,
arXiv:2004.09339 [gr-qc].


\bibitem{Gurses:2020ofy}
M.~G\"urses, T.~C.~Sisman and B.~Tekin,
Eur. Phys. J. C \textbf{80}, 647 (2020).


\bibitem{Fernandes:2020nbq}
P.~G.~S.~Fernandes, P.~Carrilho, T.~Clifton and D.~J.~Mulryne,
Phys. Rev. D \textbf{102}, 024025 (2020).


\bibitem{Kobayashi:2020wqy}
T.~Kobayashi,
JCAP \textbf{07}, 013 (2020).

\bibitem{Ma:2020ufk}
L.~Ma and H.~Lu,
arXiv:2004.14738 [gr-qc].

\bibitem{Arrechea:2020evj}
J.~Arrechea, A.~Delhom and A.~Jiménez-Cano,
arXiv:2004.12998 [gr-qc].

\bibitem{Cai:2009ua} 
R.~G.~Cai, L.~M.~Cao and N.~Ohta,
JHEP {\bf 1004}, 082 (2010).


\bibitem{Cai:2014jea}
R.~G.~Cai,
Phys.\ Lett.\ B {\bf 733}, 183 (2014).


\bibitem{Kehagias:2009is}
A.~Kehagias and K.~Sfetsos,
Phys. Lett. B \textbf{678}, 123 (2009).


\bibitem{Darwin}
C. Darwin, Proc. R. Soc. A \textbf{249}, 180 (1959).

\bibitem{Virbhadra:1999nm} 
K.~S.~Virbhadra and G.~F.~R.~Ellis,
Phys.\ Rev.\ D {\bf 62}, 084003 (2000).

\bibitem{Frittelli:1999yf} 
S.~Frittelli, T.~P.~Kling and E.~T.~Newman,
Phys.\ Rev.\ D {\bf 61}, 064021 (2000).


\bibitem{Bozza:2001xd} 
V.~Bozza, S.~Capozziello, G.~Iovane and G.~Scarpetta,
Gen.\ Rel.\ Grav.\  {\bf 33}, 1535 (2001).

\bibitem{Eiroa:2003jf} 
E.~F.~Eiroa and D.~F.~Torres,
Phys.\ Rev.\ D {\bf 69}, 063004 (2004).      

\bibitem{Whisker:2004gq} 
R.~Whisker,
Phys.\ Rev.\ D {\bf 71}, 064004 (2005).

\bibitem{Eiroa:2004gh} 
E.~F.~Eiroa,
Phys.\ Rev.\ D {\bf 71}, 083010 (2005).


\bibitem{Eiroa:2005vd}
E.~F.~Eiroa,
Braz. J. Phys. \textbf{35}, 1113-1116 (2005).
\bibitem{Li:2015vqa}
G.~Li, B.~Cao, Z.~Feng and X.~Zu,
Int.\ J.\ Theor.\ Phys.\  {\bf 54}, 3103 (2015).


\bibitem{Bhadra:2003zs} 
A.~Bhadra,
Phys.\ Rev.\ D {\bf 67}, 103009 (2003).

\bibitem{Bozza:2002zj} 
V.~Bozza,
Phys.\ Rev.\ D {\bf 66}, 103001 (2002).


\bibitem{Chen:2009eu} 
S.~b.~Chen and J.~l.~Jing,
Phys.\ Rev.\ D {\bf 80}, 024036 (2009).

\bibitem{Sarkar:2006ry} 
K.~Sarkar and A.~Bhadra,
Class.\ Quant.\ Grav.\  {\bf 23}, 6101 (2006).

\bibitem{Javed:2019qyg}
W.~Javed, R.~Babar and A.~Övgün,
Phys.\ Rev.\ D \textbf{99}, 084012 (2019).

\bibitem{Shaikh:2019itn}
R.~Shaikh, P.~Banerjee, S.~Paul and T.~Sarkar,
Phys.\ Rev.\ D \textbf{99}, 104040 (2019).


\bibitem{Eiroa:2010wm} 
E.~F.~Eiroa and C.~M.~Sendra,
Class.\ Quant.\ Grav.\  {\bf 28}, 085008 (2011).

\bibitem{Ovgun:2019wej}
A.~Övgün,
Phys.\ Rev.\ D \textbf{99}, 104075 (2019).

\bibitem{Panpanich:2019mll}
S.~Panpanich, S.~Ponglertsakul and L.~Tannukij,
Phys.\ Rev.\ D \textbf{100}, 044031 (2019).


\bibitem{Bronnikov:2018nub}
K.~Bronnikov and K.~Baleevskikh,
Grav.\ Cosmol.\  \textbf{25}, 44-49 (2019).

\bibitem{Shaikh:2018oul}
R.~Shaikh, P.~Banerjee, S.~Paul and T.~Sarkar,
Phys.\ Lett.\ B \textbf{789}, 270-275 (2019).


\bibitem {Chandrasekhar:1992}
S.~ Chandrasekhar, 
{\it The Mathematical Theory of Black Holes} (Oxford University Press, New York, 1992).

\bibitem{weinberg:1972} 
S.~ Weinberg,\textit{ Gravitation and Cosmology: Principles and Applications of the General Theory of Relativity} (New York: Wiley, 1972).


\bibitem{Bozza:2008ev} 
V.~Bozza,
Phys.\ Rev.\ D {\bf 78}, 103005 (2008).

\bibitem{Frittelli:1998hr} 
S.~Frittelli and E.~T.~Newman,
Phys.\ Rev.\ D {\bf 59}, 124001 (1999).

\bibitem{Perlick:2004zh} 
V.~Perlick, Living Rev. Relativ. 7, 9 (2004). 
\bibitem{Perlick:2003vg} 
V.~Perlick,
Phys.\ Rev.\ D {\bf 69}, 064017 (2004).


\bibitem{Bozza:2007gt} 
V.~Bozza and G.~Scarpetta,
Phys.\ Rev.\ D {\bf 76}, 083008 (2007).

\bibitem{Do:2019txf} 
T.~Do {\it et al.},
Science {\bf 365}, 664 (2019).  

\bibitem{Akiyama:2019cqa} 
K.~Akiyama {\it et al.},
Astrophys.\ J.\  {\bf 875}, L1 (2019).



\bibitem{Gibbons:2008rj} 
G.~W.~Gibbons and M.~C.~Werner,
Class.\ Quant.\ Grav.\  {\bf 25}, 235009 (2008).


\bibitem{Carmo} M. P. Do Carmo, \textit{Differential Geometry of Curves and Surfaces}, (Prentice-Hall, New Jersey, 1976).

\bibitem{Ishihara:2016vdc} 
A.~Ishihara, Y.~Suzuki, T.~Ono, T.~Kitamura and H.~Asada,
Phys.\ Rev.\ D {\bf 94},  084015 (2016).

\bibitem{Ishihara:2017}  
A.~Ishihara, Y.~Suzuki, T.~Ono and H.~Asada,
Phys.\ Rev.\ D {\bf 95}, 044017 (2017).

\bibitem{Ono:2017pie} 
T.~Ono, A.~Ishihara and H.~Asada,
Phys.\ Rev.\ D {\bf 96}, 104037 (2017).


\bibitem{Crisnejo1:2018uyn} 
G.~Crisnejo and E.~Gallo,
Phys.\ Rev.\ D {\bf 97}, 124016 (2018).

\bibitem{Ovgun:2018fnk}  
A.~\"{O}vg\"{u}n,
Phys.\ Rev.\ D {\bf 98}, 044033 (2018).

\bibitem{Ovgun:2018tua}  
A.~\"{O}vg\"{u}n, I.~Sakalli and J.~Saavedra,
JCAP {\bf 1810}, 041 (2018).


\bibitem{Javed:2019rrg}    
W.~Javed, J.~Abbas and A.~\"{O}vg\"{u}n,
Phys.\ Rev.\ D {\bf 100}, 044052 (2019).

\bibitem{Javed:2019ynm}
W.~Javed, R.~Babar and A.~\"{O}vg\"{u}n,
Phys.\ Rev.\ D {\bf 100}, 104032 (2019).

\bibitem{Javed:2019kon} 
W.~Javed, J.~Abbas and A.~\"{O}vg\"{u}n,
Eur.\ Phys.\ J.\ C {\bf 79}, 694 (2019).

\bibitem{Crisnejo:2019xtp} 
G.~Crisnejo, E.~Gallo and J.~R.~Villanueva,
Phys.\ Rev.\ D {\bf 100}, 044006 (2019).

\bibitem{Crisnejo:2018ppm}
G.~Crisnejo, E.~Gallo and A.~Rogers,
Phys.\ Rev.\ D {\bf 99}, 124001 (2019).

\bibitem{Zhu:2019ura}
T.~Zhu, Q.~Wu, M.~Jamil and K.~Jusufi,
Phys.\ Rev.\ D {\bf 100}, 044055 (2019).

\bibitem{Kumar:2019pjp}
R.~Kumar, S.~G.~Ghosh and A.~Wang,
Phys.\ Rev.\ D {\bf 100}, 124024 (2019).


\bibitem{Werner:2012rc} 
M.~C.~Werner,
Gen.\ Rel.\ Grav.\  {\bf 44}, 3047 (2012).


\bibitem{Crisnejo:2019ril}
G.~Crisnejo, E.~Gallo and K.~Jusufi,
Phys.\ Rev.\ D {\bf 100},104045 (2019).

\bibitem{Asada:2000vn} 
H.~Asada and M.~Kasai,
Prog.\ Theor.\ Phys.\  {\bf 104}, 95 (2000).
  
\end{thebibliography}
\end{document}